\documentclass[12pt]{article}
\usepackage{latexsym,graphicx,epsfig,psfrag,here}
\newcommand{\beq}{\begin{equation}}
\newcommand{\eeq}{\end{equation}}
\newcommand{\beqa}{\begin{eqnarray}}
\newcommand{\eeqa}{\end{eqnarray}}
\newcommand{\beqan}{\begin{eqnarray*}}
\newcommand{\eeqan}{\end{eqnarray*}}
\newcommand{\ba}{\begin{array}}
\newcommand{\ea}{\end{array}}
\newcommand{\ben}{\begin{enumerate}}
\newcommand{\een}{\end{enumerate}}
\newcommand{\bfl}{\begin{flushleft}}
\newcommand{\efl}{\end{flushleft}}
\newcommand{\btab}{\begin{tabular}}
\newcommand{\etab}{\end{tabular}}
\newcommand{\bit}{\begin{itemize}}
\newcommand{\eit}{\end{itemize}}
\newcommand{\bdes}{\begin{description}}
\newcommand{\edes}{\end{description}}
\newcommand{\bdm}{\begin{displaymath}}
\newcommand{\edm}{\end{displaymath}}
\newcommand{\nl}{\nonumber \\}
\newcommand{\no}{\nonumber}

\newcommand{\ve}{\varepsilon}

\newcommand{\dg}{\dagger}

\newcommand{\cL}{{\cal L}}

\newcommand{\IM}{\mbox{\rm Im}}
\newcommand{\RE}{\mbox{\rm Re}}
\newcommand{\epe}{\ve_{\pi^0\eta}} 
\newcommand{\hepph}[1]{{\tt hep-ph/#1}} 

\newcommand{\heplat}[1]{{\tt hep-lat/#1}} 
\newcommand{\nucth}[1]{{\tt nucl-th/#1}} 
\normalsize
\begin{document}
\parskip=4pt plus 1pt  
\begin{titlepage}
\begin{flushright}
UWThPh-2000-43\\
October 2000
\end{flushright}
\vspace{2cm}
\begin{center}
{\Large \bf STRONG INTERACTIONS \\[10pt] OF LIGHT FLAVOURS} \\[30pt]
{\bf Gerhard Ecker} \\[10pt]
Institut f\"ur Theoretische Physik, Universit\"at Wien \\[5pt]
Boltzmanngasse 5, A--1090 Wien, Austria \\[50pt]
\end{center}
\begin{abstract} 
\noindent After an introduction to chiral perturbation theory, the effective
field theory of the standard model at low energies, a brief overview
is given of applications to light-flavour physics. Three topics in the
strong interactions of pseudoscalar mesons are then discussed in some
detail: loops and chiral logs, pion-pion scattering and isospin
violation. 
\end{abstract}
 
\vfill
\begin{center}
Lectures given at the \\[5pt]
Advanced School on QCD 2000, Benasque, Spain \\[5pt]
July 3 - 6, 2000 \\[5pt]
To appear in the Proceedings
\end{center}
\vfill

\noindent Work supported in part by  TMR, EC-Contract No. ERBFMRX-CT980169 
(EURODA$\Phi$NE).

\end{titlepage}
\setcounter{page}{1}
\tableofcontents

\renewcommand{\thesection}{\arabic{section}}
\renewcommand{\thesubsection}{\arabic{section}.\arabic{subsection}}
\renewcommand{\theequation}{\arabic{section}.\arabic{equation}}

\setcounter{equation}{0}
\setcounter{subsection}{0}
 
\section{Chiral Perturbation Theory}
\label{sec:CHPT}

\subsection{Introduction}
The physics of light flavours at low energies can be described in the
framework of chiral perturbation theory (CHPT)
\cite{Wein79,GL84,GL85a,Leu94}. It is a typical
example of an effective field theory:
\begin{itemize} 
\item The domain of validity of CHPT is restricted to the confinement
regime of QCD, i.e., to energies below some typical scale $\Lambda$ of 
the order of $M_\rho$.
\item The matching of CHPT to the underlying standard model is a
difficult nonperturbative problem: neither CHPT nor QCD can be treated
perturbatively at $E=\Lambda$. The challenge consists in bridging the 
gap between the perturbatively accessible domains of both theories:
\begin{center}  
\begin{tabular}{ccc}
CHPT & \Large{$\leftrightarrow$}  & perturbative QCD \\[5pt]
$E < M_\rho$ & \hspace*{4cm}  & $E > 1.5$ GeV
\end{tabular}  
\end{center} 
\item The difficulty of matching the two theories is related to the 
existence of a phase transition. The degrees of freedom of the two
quantum field theories are
completely different: quarks and gluons on the one hand, hadrons, in
particular the pseudoscalar mesons as the lightest ones, on the other
hand. 
\item Symmetries are the main input for the construction of CHPT,
especially the spontaneously (and explicitly) broken chiral symmetry
of QCD. This ensures compatibility with the underlying standard model
but also leaves considerable freedom for the effective
Lagrangians. The abundance of free parameters, the so-called
low-energy constants (LECs), becomes especially acute at higher orders
in the low-energy expansion. Additional input is necessary to get
information on those constants. The most promising approaches rely on
the large-$N_c$ expansion (e.g., Refs.~\cite{KPdR99,HBB}) and on
lattice QCD (the most recent reference is \cite{HSW00}).
\end{itemize} 

\subsection{Chiral symmetry}
The starting point of CHPT is QCD in a purely theoretical setting, the
chiral limit with $N_f=$ 2 or 3 massless quarks $u,d ~[,s]$. In this
limit, the QCD Lagrangian $\cL^0_{\rm QCD}$ exhibits a global symmetry
$$
\underbrace{SU(N_f)_L \times SU(N_f)_R}_{\mbox{chiral group $G$}}
\times U(1)_V \times U(1)_A ~.
$$
$U(1)_V$ is realized as baryon number in CHPT. The axial 
$U(1)_A$ is not a symmetry at the quantum level due to the  Abelian
anomaly (see Ref.~\cite{KL00} for a recent treatment in the chiral
framework). 

There is strong evidence both from phenomenology and from theory that
chiral symmetry is spontaneously broken (see Ref.~\cite{Schlad} for
more details and references to the original work):
\begin{equation} 
G \longrightarrow H=SU(N_f)_V~,
\end{equation} 
with the conserved subgroup $H$ either isospin ($N_f=2$) or flavour
$SU(3)$ ($N_f=3$).

What is the underlying mechanism for this spontaneous breakdown ? In a
standard proof of the Goldstone theorem \cite{Goldstone} one starts from
the charge operator in a finite volume $V$, 
$Q^V=\int_V d^3x J^0(x)$ and assumes the existence of a (sufficiently 
local) operator $A$ such that the following order parameter is
nonvanishing: 
\begin{equation} 
\lim_{V\rightarrow  \infty} \langle 0|[Q^V(x^0),A]|0\rangle \neq 0 ~.
\label{eq:op}
\end{equation} 
The Goldstone theorem then tells us that there exists a massless
state $|GB\rangle$ with 
\begin{equation} 
\langle 0|J^0(0)|GB\rangle\langle GB|A|0\rangle \neq 0 ~.
\label{eq:Gth}
\end{equation} 
The relation (\ref{eq:Gth})
contains two nonvanishing matrix elements. The first one 
involves only the symmetry current and it is therefore
independent of the specific order parameter:
\begin{equation}
\langle 0|J^0(0)|GB\rangle \neq 0 \label{eq:Gme}
\end{equation}
is a necessary and sufficient condition for spontaneous
breaking. In QCD, this matrix element is determined by the meson decay
constant $F$, the chiral limit value of $F_\pi=92.4$ MeV.

Which then is the order parameter of chiral symmetry
breaking in QCD ? Since the spontaneously broken currents are axial, 
$A$ in (\ref{eq:op}) must be a pseudoscalar, colour singlet
operator. The unique choice for local
operators in QCD with lowest operator dimension three
is\footnote{The $\lambda_i$ are the generators of $SU(N_f)_V$
in the fundamental representation.}
\begin{equation} 
A_i = \overline{q}\gamma_5 \lambda_i q
\end{equation} 
with
\begin{equation} 
\left[Q^i_A,A_j\right]=-\displaystyle\frac{1}{2} \overline{q}
\{\lambda_i,\lambda_j\} q ~.
\end{equation} 
If the vacuum is invariant under $SU(N_f)_V$, 
\begin{equation} 
\langle 0|\overline{u}u|0\rangle=\langle 0|\overline{d}d|0\rangle
\left[=\langle 0|\overline{s}s|0\rangle\right]~.
\end{equation} 
Therefore, a nonvanishing quark condensate
\begin{equation} 
\langle 0|\overline{q}q|0\rangle \neq 0 \label{eq:qcond}
\end{equation} 
is sufficient for spontaneous chiral symmetry
breaking, but certainly not necessary. Increasing the
operator dimension, the next candidate is the so--called mixed
condensate of dimension five ($\langle 0|\overline{q}\sigma_{\mu\nu}
\lambda_\alpha q G^{\alpha\mu\nu} |0\rangle \neq 0$),
and there are many more possibilities for operator dimensions 
$\ge 6$. Although all order parameters are in principle equally good for
triggering the Goldstone mechanism one may expect a special role for 
the quark condensate as the dominant order 
parameter of spontaneous chiral symmetry breaking. 

It has recently been argued that the answer could depend crucially on
$N_f$ because there are indications for $\langle
0|\overline{q}q|0\rangle$ decreasing with $N_f$ \cite{Mouss00,DGS99}.
As I will point out later on, the Gell-Mann--Okubo mass formula for
the pseudoscalar mesons supports a large quark condensate for
$N_f=3$. If the condensate indeed increases for smaller $N_f$, the
dominance should be all the more pronounced in the two-flavour
case. As a matter of fact, recent developments in $\pi\pi$ scattering 
(see Sec.~\ref{sec:pipi}) leave little doubt about the dominant order
parameter for $N_f=2$.

\subsection{Chiral Lagrangians}
Spontaneously broken symmetries are realized nonlinearly on the
Goldstone boson fields of which there are dim $G/H = N_f^2 -1$ in the
case of chiral symmetry: the pions for $N_f=2$, with kaons and the 
$\eta$ meson in addition for $N_f=3$. This nonlinear realization is
implemented by matrix fields (usually in the fundamental
representation) $u_L(\phi),u_R(\phi)$ that parametrize the coset space
$G/H$ in terms of the Goldstone fields $\phi$. Chiral transformations
$g=(g_L,g_R)\in G$ are realized as
\begin{equation} 
u_A(\phi)\stackrel{g}{\longrightarrow}  
g_A u_A(\phi) h(g,\phi)^{-1} \qquad (A=L,R)
\end{equation} 
with a so-called compensator (field) $h(g,\phi) \in SU(N_f)_V$. Its
dependence on the fields $\phi$ is a characteristic feature of the
nonlinear realization. 

In the purely mesonic sector, the Goldstone fields are usually
parametrized in terms of another matrix field $U(\phi)$ that
transforms linearly under chiral transformations:
\begin{equation} 
U(\phi):=u_R(\phi)u_L(\phi)^\dg \stackrel{g}{\longrightarrow}  
g_R U(\phi) g_L^{-1}~.
\end{equation}
The nonlinear realization of $G$ on $\phi$ implies that the matrix
$U(\phi)$ cannot be a polynomial function of $\phi$. One standard
choice is the exponential parametrization with
\begin{equation} 
u_R(\phi)=u_L(\phi)^\dg:=u(\phi)=\exp{i \lambda_a \phi^a /2 F}
\end{equation} 
$$
\longrightarrow \qquad U(\phi)=u(\phi)^2
$$
where $F$ is defined by the Goldstone matrix element :
\begin{equation} 
\langle 0|\overline{q(x)}\gamma^\mu\gamma_5\frac{\lambda_a}{2} 
q(x)|\phi_b(p)\rangle=i\delta_{ab} F p^\mu e^{-ipx}~.
\label{eq:F}
\end{equation} 
As chiral Lagrangians are built from the matrix fields $u_{L,R}$ or
$U$ it is clear that the corresponding quantum field theories are
generically nonrenormalizable.

It is now time to take leave from the theorist's world of chiral
symmetry and admit that there is no chiral symmetry in nature. In the
standard model, chiral symmetry is explicitly broken in two different 
ways:
\begin{enumerate} 
\item[i.] Nonvanishing quark masses: this is expected to be a small
deviation from the chiral limit for two flavours but less so for
$N_f=3$. 
\item[ii.] Electroweak interactions: these can be taken into account
perturbatively in $\alpha$ and $G_F$ (more about this in 
Sec.~\ref{sec:app}).
\end{enumerate} 

The main assumption underlying CHPT is that an expansion
around the chiral limit is a meaningful approximation. Even
neglecting the electroweak interactions, i.e., for $\alpha=G_F=0$,
CHPT is from the outset based on a two-fold expansion, both in the
momenta of pseudoscalar mesons and in the quark masses. For the
effective chiral Lagrangian $\cL_{\rm eff}$, this implies
\begin{equation} 
\cL_{\rm eff} = \sum_{i,j} \cL_{ij}~, \qquad 
\cL_{ij} = O(\partial^i m^j_q)
\end{equation} 
where $\partial$ stands for a derivative.

The two expansions can be related to each other by making use of the
relation between meson and quark masses:
\begin{eqnarray}
\label{eq:B}
M_M^2 & \sim & B m_q + O(m_q^2) \\
B & = & - \langle 0|\overline{u}u|0\rangle / F^2 ~.\no
\end{eqnarray} 
Depending on the value of the quark condensate, two different schemes 
have been considered.
\begin{center} 
A. Standard CHPT
\end{center} 
This is the original scheme \cite{Wein79,GL84,GL85a} where the terms
linear in the quark masses are assumed to dominate the meson masses in
(\ref{eq:B}). This corresponds to a value of $B(\nu=1~\rm GeV)\sim$
1.4 GeV ($\nu$ is the QCD renormalization scale) and gives rise to the 
standard values of light quark mass ratios.
It also implies the Gell-Mann--Okubo mass
formula $3M^2_{\eta_8} = 4 M_K^2 - M_\pi^2$ at lowest order in
the chiral expansion. The standard chiral counting is 
\begin{equation} 
m_q=O(M^2)= O(p^2)~,
\end{equation}
implying in turn 
\begin{equation} 
\cL_{\rm eff}  = \sum_n \cL_n~, \qquad 
\cL_n = \sum_{i+2j=n} \cL_{ij} ~.
\end{equation}

\begin{center} 
\item[B.] Generalized CHPT
\end{center} 
The proponents of the second scheme (\cite{Stern97} and references
therein) allow for the possibility that $B$ is much smaller, e.g.,
$ B(\nu = 1 {\rm ~GeV})=O(F_\pi)$. In this case,
\begin{equation}  
M_M^2 \sim  O(m_q^2)
\end{equation} 
and the light quark mass ratios would be quite different from the
standard values. The more natural chiral counting is $m_q= O(p)$ now
and the (same) effective Lagrangian is reordered, with more terms
catalogued at lower orders than in the standard expansion. The obvious
drawback is that there are more unknown LECs at any given order. 
For instance, the Gell-Mann--Okubo mass formula is not a consequence
of Generalized CHPT. It is
therefore rather comforting that there is at present no compelling evidence
for scheme B. On the other hand, it is difficult to prove the validity
of scheme
A, especially for three light flavours. I will come back in
Sec.~\ref{sec:pipi} to the situation in the two-flavour case where 
evidence for the standard procedure has been mounting  recently.

To construct the effective chiral Lagrangian(s), we follow the
procedure of Gasser and Leutwyler \cite{GL84,GL85a} by coupling external
matrix fields $v_\mu,a_\mu,s,p$ to the quarks:
\begin{equation} 
\cL^0_{\rm QCD} \to \cL^0_{\rm QCD} + \overline q 
\gamma^\mu(v_\mu + 
a_\mu \gamma_5)q - \overline q (s - ip \gamma_5)q ~.
\end{equation} 
Chiral symmetry is promoted to a local symmetry in this way. QCD Green
functions and amplitudes at low energies can then be calculated from a
quantum field theory with the most general effective Lagrangian that
respects this local chiral symmetry \cite{Leu94}.

In the standard scheme, this effective chiral Lagrangian for the strong
interactions of pseudoscalar mesons takes the general form
\begin{equation} 
\cL_{\rm eff}  = \cL_2 + \cL_4 + \cL_6 + \dots 
\end{equation} 
The lowest-order Lagrangian $\cL_2$ of $O(p^2)$ is given by
\begin{equation} 
\cL_2  = \frac{F^2}{4} \langle D_\mu U D^\mu U^\dagger + 
\chi U^\dagger + \chi^\dagger  U \rangle 
\end{equation} 
with a gauge-covariant derivative $D_\mu U = \partial_\mu U - 
i (v_\mu + a_\mu) U + i U (v_\mu - a_\mu)$ and with
$\chi = 2B(s + ip)$. The two free parameters $F,B$ are related to the
pion decay constant and to the quark condensate, respectively:
\begin{eqnarray} 
F_\pi &=& F \left[1+O(m_q)\right]  \\
\langle 0| \overline{u}u|0\rangle &=& - F^2  B
\left[1+O(m_q)\right]~. \no
\end{eqnarray} 

CHPT at lowest order amounts to the calculation of Green
functions and amplitudes with $\cL_2$ at tree level where one sets
all external fields to zero at the end except for
\begin{equation} 
s(x)={\rm diag}(m_u,m_d [,m_s])
\end{equation} 
to account for the explicit chiral symmetry breaking through the quark
masses. The results are equivalent to the current algebra amplitudes
of the sixties. Amplitudes depend only on $F_\pi$ and $M_M$, e.g., the
Weinberg amplitude \cite{Wein66} for pion-pion scattering:
\begin{equation} 
A_2(s,t,u) = \frac{s-M_\pi^2}{F_\pi^2} ~.
\label{eq:Wein}
\end{equation} 
It is remarkable that one seems to get an absolute prediction 
from pure symmetry only! The skeptic will soon realize that it is
in fact a relation between a four-point function and the two-point
function (\ref{eq:F}), which is only possible for a nonlinearly
realized symmetry.

At next-to-leading order in the chiral expansion, 
the chiral Lagrangian $\cL_4$ contains 
7 (10) measurable LECs for $N_f=$ 2 (3) \cite{GL84,GL85a}. At this
point, we have to take the quantum field theory character of
CHPT serious. Since $\cL_{\rm eff}$ is hermitian tree amplitudes are
real. On
the other hand,  unitarity and analyticity demand complex amplitudes
in general. For instance, the partial-wave amplitudes in $\pi\pi$
scattering (cf. Sec.~\ref{sec:pipi}) satisfy 
\begin{equation} 
\IM t^I_l(s) \ge 
(1-\displaystyle\frac{4 M_\pi^2}{s})^{\frac{1}{2}} 
|t^I_l(s)|^2~.
\end{equation}
The Weinberg amplitude (\ref{eq:Wein}) produces (real)
partial waves $t^I_l(s)$ of $O(p^2)$ (for $l \le 1$). Thus, the 
scattering amplitude must have a nonvanishing imaginary part, starting 
at $O(p^4)$. 

A systematic low-energy expansion to $O(p^n)$ requires therefore a
concurrent loop expansion for $n > 2$. Since CHPT loop amplitudes are at 
least as divergent as in renormalizable quantum field theories, the 
theory has to be regularized {\bf and} renormalized. Renormalization is
essential for getting cutoff independent results. The procedure
amounts to absorbing divergences by the coupling constants in 
$\cL_4, \cL_6, \dots$ rendering the observable LECs scale dependent at
the same time. This scale dependence is by construction always compensated
by the scale dependence of loop amplitudes.

The LECs contain the effect of all those (heavy) degrees of freedom
that do not appear as explicit fields in $\cL_{\rm eff}$. For
instance, the effective Lagrangian for $N_f=2$ contains only
pions. Kaons and the $\eta$ enter only via the coupling
constants. For $N_f=3$, the dominant degrees of freedom governing the
size of LECs are the meson resonances \cite{EGPR89}.
 
\setcounter{equation}{0}
\setcounter{subsection}{0}
 
\section{Survey of Applications}
\label{sec:app}
Before discussing a few selected examples in some detail, I want to
give a brief overview of the rapidly growing field of
applications of CHPT. One way to summarize the current status is to
present the effective chiral Lagrangian of the standard model in Table
\ref{tab:EFTSM}.

\renewcommand{\arraystretch}{1.3}
\begin{table}[ht]
\begin{center}
\caption{The effective chiral Lagrangian of the Standard Model}
\label{tab:EFTSM}
\vspace{.5cm}
\begin{tabular}{|l|c|} 
\hline
&  \\
\hspace{1cm} ${\cal L}_{\rm chiral\; order}$ 
~($\#$ of LECs)  &  loop  ~order \\[8pt] 
\hline 
&  \\
${\cal L}_{p^2}(2)$+${\cal L}_{p^4}^{\rm odd}(0)$+
${\cal L}_{G_Fp^2}^{\Delta S=1}(2)$ +${\cal L}_{e^2p^0}^{\rm em}(1)$
+${\cal L}_{G_8e^2p^0}^{\rm emweak}(1)$  & $L=0$\\[8pt]
~+~${\cal L}_{p}^{\pi N}(1)$+${\cal L}_{p^2}^{\pi N}(7)$
+${\cal L}_{G_8p^0}^{MB,\Delta S=1}(2)$+
${\cal L}_{G_8p}^{MB,\Delta S=1}(8)$ 
+ ${\cal L}_{e^2p^0}^{\pi N,{\rm em}}(3)$ & \\[15pt]
~+~$\underline{{\cal L}_{p^4}^{\rm even}(10)}$+
$\underline{{\cal L}_{p^6}^{\rm odd}(32)}$
+$\underline{{\cal L}_{G_8p^4}^{\Delta S=1}(22)}$
+$\underline{{\cal L}_{e^2p^2}^{\rm em}(14)}$ +
$\underline{{\cal L}_{G_8e^2p^2}^{\rm emweak}(14)}$ & $L=1$\\[8pt]
~+~$\underline{{\cal L}_{e^2p}^{\rm leptons}(5)}$ &\\[8pt]   
~+~$\underline{{\cal L}_{p^3}^{\pi N}(23)}$+
$\underline{{\cal L}_{p^4}^{\pi N}(114)}$
+ $\underline{{\cal L}_{G_8p^2}^{MB,\Delta S=1}(?)}$ 
+$\underline{{\cal L}_{e^2p}^{\pi N,{\rm em}}(8)}$ 
 &  \\[15pt]
~+~$\underline{{\cal L}_{p^6}^{\rm even}(90)}$  & $L=2$ \\[8pt] 
\hline
\end{tabular}
\end{center}
\end{table}

The various Lagrangians are ordered by their chiral dimension, with the
number of independent LECs shown in brackets. Except for the pieces
with superscript $\pi N$, the numbers refer to $N_f=3$. The fully
renormalized Lagrangians are underlined. This means that the
divergence structure of the corresponding loop functionals is
explicitly known in a process independent manner.

The Table indicates that CHPT is not restricted to the pseudoscalar
mesons only (Lagrangians ${\cal L}_{p^2}(2)$, 
${\cal L}_{p^4}^{\rm odd}(0)$, ${\cal L}_{p^4}^{\rm
even}(10)$, ${\cal L}_{p^6}^{\rm odd}(32)$ and
${\cal L}_{p^6}^{\rm even}(90)$).
The inclusion of baryons is straightforward but there are
substantial differences to the purely mesonic case:
\begin{itemize}
\item Baryons are not Goldstone modes: there are no ``soft'' 
baryons. Their interactions are less constrained by chiral symmetry.
\item In the presence of baryons, the effective Lagrangian has parts
of every integer chiral order even in the standard scheme:
\begin{equation} 
\cL_{\rm eff}  = \sum_{n=1,2,3,\dots} \cL_n ~.
\end{equation} 
Compared to the purely mesonic case with $n=2,4,6,\dots$, the chiral
expansion progresses much more slowly. 
\item Baryon resonances are often closer to threshold than in the case
of mesons ($\Delta$ vs. $\rho$). This limits the domain of validity of
the chiral expansion. 
\item The baryon mass $m$ complicates the chiral counting because it
does not vanish in the chiral limit. The traditional method is called
Heavy Baryon CHPT \cite{JM91} and it amounts to shifting the baryon
mass from the propagators to the vertices of an effective
Lagrangian. The method is systematic and straightforward but it does
not converge in some kinematic configurations. Therefore, an
alternative called Relativistic Baryon CHPT has recently been put
forward \cite{BL99} that is manifestly Lorentz invariant at every step
and does not have the deficiency of HBCHPT. Loop calculations and the 
regularization procedure are somewhat more involved. Both approaches
have been applied to a variety of processes, albeit most of them only
in HBCHPT so far. The corresponding Lagrangians in Table
\ref{tab:EFTSM} are ${\cal L}_{p}^{\pi N}(1)$, 
${\cal L}_{p^2}^{\pi N}(7)$, ${\cal L}_{p^3}^{\pi N}(23)$
and ${\cal L}_{p^4}^{\pi N}(114)$.
\item Table \ref{tab:EFTSM} is restricted to single-baryon
processes. It does not account for applications of chiral methods
in nuclear physics such as in nucleon-nucleon
scattering (see Ref.~\cite{nucph} for a review of recent developments 
in this field).
\end{itemize} 

Other heavy fields can also be included in the effective Lagrangians,
although at the expense of possible double counting in some cases,
e.g., the meson and baryon resonances.

So far, I have only considered strong interactions. Electroweak
interactions can be incorporated perturbatively in $G_F$ and
$\alpha$. The previous method of constructing effective chiral
Lagrangians has to be extended in this case. To describe the
nonleptonic weak interactions of mesons treated by Hans Bijnens in his
lectures \cite{HBB}, one first has to integrate out the heavy fields
$W,t,b,c$ to obtain an effective Hamiltonian for $\Delta S=1$
nonleptonic weak interactions. The task is then to construct the most
general effective chiral Lagrangian with the same chiral transformation
properties as this effective Hamiltonian.
The corresponding Lagrangians in Table 
\ref{tab:EFTSM} are ${\cal L}_{G_Fp^2}^{\Delta S=1}(2)$,
${\cal L}_{G_8p^4}^{\Delta S=1}(22)$ for mesons only and 
${\cal L}_{G_8p^0}^{MB,\Delta S=1}(2)$,
${\cal L}_{G_8p}^{MB,\Delta S=1}(8)$, 
${\cal L}_{G_8p^2}^{MB,\Delta S=1}(?)$ 
for baryons and mesons.

Inclusion of the electromagnetic interactions is still a little more
involved, being nonlocal at low energies. One introduces the dynamical
photon with the proper kinetic term (and gauge fixing) and an
additional chiral Lagrangian that transforms like a product of two
electromagnetic currents. The corresponding Lagrangians are 
${\cal L}_{e^2p^0}^{\rm em}(1)$, ${\cal L}_{e^2p^2}^{\rm em}(14)$ 
(mesons) and ${\cal L}_{e^2p^0}^{\pi N,{\rm em}}(3)$, 
${\cal L}_{e^2p}^{\pi N,{\rm em}}(8)$ (baryons and mesons). The
combination of nonleptonic weak and electromagnetic interactions
requires still another set of Lagrangians that have only been considered 
for mesons so far :  ${\cal L}_{G_8e^2p^0}^{\rm emweak}(1)$, 
${\cal L}_{G_8e^2p^2}^{\rm emweak}(14)$.

Finally, leptons can also be incorporated as explicit fields in the
effective Lagrangian, up to now again with mesons only:  
${\cal L}_{e^2p}^{\rm leptons}(5)$.

Instead of trying to give proper credit to all the work
related to the effective Lagrangians in Table \ref{tab:EFTSM},
I refer to some recent workshops from where up-to-date information 
can be recovered \cite{Mainz,Honnef,Cebaf}.

\setcounter{equation}{0}
\setcounter{subsection}{0}
 
\section{Loops and Chiral Logs}
\label{sec:loops}
\subsection{Loop expansion}

Successive orders in the chiral expansion can be characterized by 
the chiral dimension $D_L$ defined as the degree of homogeneity of
amplitudes in
external momenta and meson masses. In the meson sector with the
standard chiral counting, a generic $L$-loop diagram with $N_n$
vertices from the Lagrangian $\cL_n ~(n=2,4,6,\dots)$ and $I$ internal
lines has $D_L = 4L + \sum_{n \ge 2} nN_n - 2I$. The topological relation
$L=I-\sum_{n} N_n +1$ for connected diagrams leads to the final
expression \cite{Wein79}
\begin{equation} 
D_L = 2L + 2 + \sum_{n \ge 4} (n-2)N_n \ge 2L + 2~.
\label{eq:cD}
\end{equation} 

As $D_L$ increases with $L$ while the physical dimension of a given
amplitude remains fixed, each loop comes with a factor 
$1/(4\pi F)^2$ so that the chiral expansion is really an expansion in
\begin{equation} 
\frac{p^2}{(4\pi F_\pi)^2}= 0.18 \frac{p^2}{M_K^2} ~.
\end{equation} 
Corrections of the order of 20\% are therefore to be expected in the
three-flavour case. For $N_f=2$, one may limit the expansion to
smaller momenta with correspondingly improved precision.

\begin{figure}[ht]
\centerline{\epsfig{file=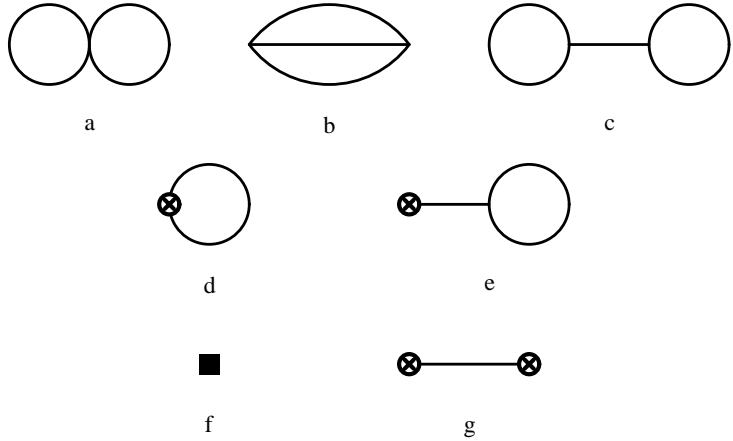,height=6cm}}
\caption{Skeleton diagrams of $O(p^6)$. Propagators and vertices carry 
the full tree structure associated with the lowest-order Lagrangian
$\cL_2$. Normal vertices are from
$\cL_2$, crossed circles and the full square denote vertices from 
$\cL_4$ and $\cL_6$, respectively.}
\label{fig:p6diag}
\end{figure}

The state of the art for the strong interactions of mesons is
$O(p^6)$. The possible values of $L$ and $N_n$ ($n=4,6$) in 
(\ref{eq:cD}) for $D_L=6$ are shown graphically in Fig.~\ref{fig:p6diag} 
as so-called skeleton diagrams. For actual calculations to $O(p^6)$, I
refer again to the workshops listed in Refs.~\cite{Mainz,Honnef,Cebaf}.

The renormalization procedure is best carried out in a
process independent way via the divergent part of the generating
functional of Green functions. Such an approach has several
advantages:
\begin{itemize} 
\item It provides a nontrivial check for all explicit loop calculations
for specific processes. 
\item It produces renormalization group equations for the
renormalized, scale-dependent LECs.
\item As an important byproduct, one gets the leading infrared 
singularities for free, the so-called chiral logs.
\end{itemize} 

\subsection{Renormalization at $O(p^4)$}
In dimensional regularization, divergences appear in the 
combination
\begin{equation} 
\frac{\mu^{d-4}}{(4\pi)^2} 
\left[\frac{1}{d-4} + \frac{1}{2}\ln{M^2/\mu^2}+\dots \right]
\end{equation} 
that is independent of the arbitrary renormalization scale $\mu$. $M$
denotes a typical scale, e.g., $M_\pi$ ($M_K$) for $N_f=$ 2
(3). Renormalization consists in splitting the coupling constants
$L_i$ of the Lagrangian $\cL_4$ in an analogous fashion :
\begin{equation} 
L_i=\mu^{d-4}\left[\frac{\Gamma_i}{(4\pi)^2(d-4)}+
\underbrace{L_i^r(\mu)}_{{\rm ren. ~LEC }}+\dots  \right]~.
\end{equation} 
The coefficients $\Gamma_i$ are chosen such that the divergences of
one-loop Green functions are cancelled by the tree-level contributions
from $\cL_4$. As a consequence, physical amplitudes depend on the 
scale-independent combinations
\begin{equation} 
L_i^r(\mu) - \frac{1}{2}\Gamma_i l
\end{equation} 
with the chiral log 
$$ 
l=\frac{1}{(4\pi)^2}\ln{M^2/\mu^2}~.
$$
Scale independence of these combinations implies renormalization
group equations for the measurable LECs $L_i^r(\mu)$~:
\begin{eqnarray} 
\mu\frac{d L_i^r(\mu)}{d\mu} &=& -\frac{\Gamma_i}{(4\pi)^2} \\
\to\qquad L_i^r(\mu_2) &=& L_i^r(\mu_1) + 
\frac{\Gamma_i}{16\pi^2} \ln\frac{\mu_1}{\mu_2} ~.\no
\end{eqnarray} 

The phenomenological values of the LECs for $N_f=3$ are shown in 
Table \ref{tab:LEC}. Some of them have recently been 
reevaluated on the basis of an $O(p^6)$ analysis \cite{ABT00}.

\renewcommand{\arraystretch}{1.1}
\begin{table}[ht]
\begin{center}
\caption{ Phenomenological values of $L^r_i(M_\rho)$}
\label{tab:LEC}
\vspace*{.2cm}
\begin{tabular}{|c||r|r||l|}  \hline
&\multicolumn{2}{c||}{} & \\
i &\multicolumn{2}{c||} {\mbox{  } $L^r_i(M_\rho) \times 10^3$ 
\mbox{  }} & \\[5pt]
 &\multicolumn{2}{c||} { 1995 \hspace*{2cm}
\mbox{    } 2000} & \hspace*{.3cm} 
main source \\
 &\multicolumn{2}{r||}{ \mbox{}
 (Amoros et al. \cite{ABT00})} &  \\[5pt]
\hline
  1  & \hspace*{.3cm} 0.4 $\pm$ 0.3 \hspace*{.3cm}& \hspace*{.3cm} 0.52 $\pm$ 0.23 \hspace*{.3cm}& $K_{e4},
(\pi\pi\to\pi\pi)$  \\
  2  & \hspace*{.3cm} 1.35 $\pm$ 0.3 \hspace*{.3cm} &  
\hspace*{.3cm} 0.72 $\pm$ 0.24 \hspace*{.3cm} & $K_{e4},
(\pi\pi\to\pi\pi)$ \\
  3  & \hspace*{.3cm} $-$3.5 $\pm$ 1.1 \hspace*{.3cm}& \hspace*{.3cm} $-$2.70 $\pm$ 0.99 \hspace*{.3cm}& $K_{e4},
(\pi\pi\to\pi\pi)$    \\
  4  & \hspace*{.3cm} $-$0.3 $\pm$ 0.5 \hspace*{.3cm} &  & Zweig rule   \\
  5  & \hspace*{.3cm} 1.4 $\pm$ 0.5  \hspace*{.3cm} & \hspace*{.3cm}  0.65 $\pm$ 0.12 \hspace*{.3cm} &$F_K/F_\pi$  \\
  6  & \hspace*{.3cm} $-$0.2 $\pm$ 0.3 \hspace*{.3cm} & \hspace*{.3cm}   & Zweig rule  \\
  7  & \hspace*{.3cm} $-$0.4 $\pm$ 0.2 \hspace*{.3cm} & \hspace*{.3cm}  $-$0.26 $\pm$ 0.15 \hspace*{.3cm} &
Gell-Mann--Okubo, \\
     &     &   & $L_5,L_8$       \\
  8  & \hspace*{.3cm} 0.9 $\pm$ 0.3 \hspace*{.3cm} & \hspace*{.3cm}  0.47 $\pm$ 0.18 \hspace*{.3cm} &$M_{K^0}-M_{K^+},L_5$\\
 9  & \hspace*{.3cm} 6.9 $\pm$ 0.7 \hspace*{.3cm} & \hspace*{.3cm}  \hspace*{.3cm} &$\langle r^2\rangle^\pi_V$  \\
 10  & \hspace*{.3cm} $-$5.5 $\pm$ 0.7\hspace*{.3cm} &  & $\pi \rightarrow e \nu\gamma$   \\
\hline
\end{tabular}
\end{center}
\end{table}

\subsection{Renormalization at $O(p^6)$}
The procedure here is similar, albeit more complicated than before due
to the presence of loop diagrams with $L \le 2$. Let us first have a
look at the reducible diagrams c,e,g in Fig.~\ref{fig:p6diag}. One
expects that the sum of these diagrams is finite (and
scale independent) because the one-loop renormalization has already
been carried out. This is in fact true, but only with an additional
technical requirement that would not occur in renormalizable quantum 
field theories: the Lagrangian $\cL_4$ has to be chosen appropriately
\cite{BCE3}.

Turning to the irreducible diagrams a,b,d in
Fig.~\ref{fig:p6diag}, general theorems of renormalization theory tell
us that the divergences must be polynomials in masses and external
momenta of $O(p^6)$. This is a highly nontrivial constraint on the
procedure because each type of diagrams a,b,d separately has in
addition nonlocal divergences. For instance, diagrams a and d involve
the Green function $G(x,x)$ (and derivatives):
\begin{equation} 
G(x,x)=\frac{2\mu^{d-4}}{(4\pi)^2} 
\left(\frac{1}{d-4} + \frac{1}{2}\ln{M^2/\mu^2}\right)a_1(x,x)+
\underbrace{\overline G(x,x)}_{\rm finite, nonlocal}~,
\end{equation} 
with $a_1(x,x)$ a (local) Seeley-DeWitt coefficient. Therefore,
diagrams a,b and d separately have nonlocal divergences of the form
\begin{equation} 
\frac{1}{d-4}\overline G(x,x) \quad {\rm and} \quad
\frac{1}{d-4}\ln{M^2/\mu^2} 
\end{equation} 
that have to cancel in the sum. As already emphasized, this
requirement is a very efficient check on the correctness of the
renormalization procedure \cite{BCE3}. 

The remaining divergences are indeed polynomials in masses and
derivatives (momenta) of $O(p^6)$. Those divergences are canceled by
the divergent parts of the LECs of $O(p^6)$ in the tree-level
amplitudes characterized by the vertex f in Fig.~\ref{fig:p6diag}. The
corresponding effective Lagrangian has the general form \cite{BCE2}
\begin{equation} 
\cL_6  = \sum_{i=1}^{53(90)}\underbrace{C_i}_{\rm coeffs.} 
\underbrace{O_i}_{\rm monomials} \quad {\rm for} \quad N_f=2(3)~.
\end{equation} 

The sum of all diagrams in Fig.~\ref{fig:p6diag} and therefore the
complete generating functional of Green functions of $O(p^6)$ is then
finite and scale independent with renormalized LECs  $C_i^r(\mu)$.
These LECs satisfy renormalization group equations of the form
\cite{BCE3}
\begin{equation} 
\mu\frac{d C_i^r(\mu)}{d\mu}=
\frac{1}{(4\pi)^2}\left[2\hat{\Gamma}^{(1)}_i+
\hat{\Gamma}^{(L)}_i(\mu)
\right] 
\end{equation} 
where the $\hat{\Gamma}^{(1)}_i$ are constants and 
$\hat{\Gamma}^{(L)}_i(\mu)$ are linear combinations of the LECs 
$L_i^r(\mu)$  of $O(p^4)$.

\subsection{Chiral logs} 
Chiral logs are due to the pseudo-Goldstone nature of the
pseudoscalar mesons. For reasonable choices of $M$ and $\mu$ in
\begin{equation} 
l=\frac{1}{(4\pi)^2}\ln{M^2/\mu^2}
\label{eq:log}
\end{equation} 
the contributions of chiral logs are often numerically important or
even dominant, e.g., in $\pi\pi$ scattering (cf. Sec.~\ref{sec:pipi}).
On the other hand, physical amplitudes are of course independent of
the arbitrary scale $\mu$ so why not choose $\mu=M$ and let the chiral
logs disappear?

The answer is that the choice $\mu=M$, especially for $M=M_\pi$ in the 
two-flavour case, generates unnaturally large LECs $L_i^r(M)$ (and
likewise for LECs of higher orders) because infrared effects are
in this way shifted into the coupling constants that originate 
actually from the short-distance part of the theory. In other words, the
natural size of renormalized LECs can only be understood in terms of
higher-mass states such as meson resonances \cite{EGPR89} for 
$\mu \simeq M_\rho$. 

At $O(p^6)$, the leading infrared singularities are squares of $l$ in 
(\ref{eq:log}), the so-called double chiral logs. They appear in the
following combinations in the two types of diagrams in 
Fig.~\ref{fig:p6diag}:\\[.1cm] 
\begin{tabular}{lc} 
irreducible~: &  $4 L_i^r(\mu) l - \Gamma_i l^2 := k_i$  \\
reducible~: &  $[L_i^r(\mu) - \frac{1}{2}\Gamma_i l]
[L_j^r(\mu) - \frac{1}{2} \Gamma_j l] =
L_i^r(\mu)L_j^r(\mu)-\frac{1}{8}(\Gamma_i k_j+ \Gamma_j k_i)$ ~.
\end{tabular} 

Therefore, the full dependence on $l^2, l L_i^r, L_i^r L_j^r$ can be
expressed in terms of  $k_i$ and $L_i^r L_j^r$ (generalized double-log
approximation \cite{BCE1}). Moreover, the coefficients of these terms
can be calculated from diagrams with $L \le 1$ only \cite{Wein79}.

It need hardly be emphasized that the double-log approximation 
yields at best an indication of the size of $O(p^6)$ corrections. It
is by no means a substitute for a full calculation. In fact, for the
case of pion-pion scattering chiral logs give the dominant
corrections for some observables.
As an example in the three-flavour case, I consider the ratio
$F_K/F_\pi = 1.22 \pm 0.01$ that receives sizable corrections from
double chiral logs of the order of $6 \div 12\%$ \cite{BCE1}. The 
$O(p^4)$ result 
for $F_K/F_\pi -1$ was used \cite{GL85a} to fix the LEC $L_5^r$ 
as $L_5^r(M_\rho)=(1.4 \pm 0.5)\times 10^{-3}$. A recent fit on the
basis of a full $O(p^6)$ calculation \cite{ABT00} has instead led to a
value (cf. Table \ref{tab:LEC}) 
$$
L_5^r(M_\rho)=(0.65 \pm 0.12)\times 10^{-3}~,
$$
confirming the trend indicated by the generalized double-log
approximation.

\setcounter{equation}{0}
\setcounter{subsection}{0}
 
\section{Pion-Pion Scattering}
\label{sec:pipi}
Pion-pion scattering is {\bf the} fundamental scattering process for
CHPT with $N_f=2$. The scattering amplitude near threshold is
sensitive to the mechanism of spontaneous chiral symmetry breaking,
i.e., to the size of the quark condensate.

After a long break without much experimental activity, the situation
has now improved significantly. First results from a $K_{e4}$
experiment at Brookhaven are already available to extract
pion-pion phase shifts due to final-state interactions, with more to
come from KLOE at  DA$\Phi$NE and from NA48 at CERN. In addition, the 
ambitious DIRAC experiment is well under way at CERN
to measure a combination of $S$-wave scattering lengths in
pionium, electromagnetically bound $\pi^+\pi^-$ states.

In the isospin limit $m_u=m_d$, the scattering amplitude is
determined by one scalar function $A(s,t,u)$ of the
Mandelstam variables. In terms of this function, one can
construct amplitudes with definite isospin ($I=0,1,2$) in the
$s$--channel. Partial-wave amplitudes $t_l^I(s)$ are parametrized
in terms of phase shifts $\delta_l^I(s)$ in the elastic region
$4M_\pi^2 \leq s \leq 16M_\pi^2$.
The behaviour of partial waves near threshold is of the form
\begin{equation}
\RE t_l^I(s)=q^{2l}\{a_l^I +q^2 b_l^I +O(q^4)\}~,
\label{eq:effr}
\end{equation}
with $q$ the center-of-mass momentum.  The quantities $a_l^I$ and
$b_l^I$ are referred to as scattering lengths and slope parameters,
respectively.

\subsection{Chiral expansion}
The low-energy expansion for $\pi\pi$ scattering has been
carried through to $O(p^6)$. At lowest order, the scattering
amplitude (\ref{eq:Wein}) gives rise to partial waves with $l \le 1$
only. At the same order in the standard scheme, the quark mass ratios 
are fixed in terms of meson mass ratios, in particular
\begin{equation}
r:=\frac{m_s}{\hat{m}} =
r_2:=\frac{2 M_K^2}{M_\pi^2}-1 \simeq 26 
\end{equation}
with $2 \hat{m} := m_u+m_d$.

The situation is different in the generalized scenario because there
are more parameters already at lowest order. The ratio $r$ can vary in
the range $6 \le r \le r_2$ and the scattering amplitude can be
written in the form \cite{SSF93}
\begin{equation} 
A_2^{\rm GCHPT}(s,t,u)=\frac{s-\frac{4}{3}M_\pi^2}{F_\pi^2}
+ \alpha \frac{M_\pi^2}{3F_\pi^2}
\end{equation} 
with
$$
\alpha= 1 + \frac{6(r_2-r)}{r^2-1}~, \qquad
\alpha \ge 1~.
$$
The amplitude is correlated with the quark mass ratio $r$.
Especially the $S$-waves are very sensitive to $\alpha$: the
standard lowest-order value of the scattering length $a_0^0=0.16$ 
for $\alpha=1$ ($r=r_2$) moves to $a_0^0=0.26$ for a typical value of 
$\alpha\simeq 2$ ($r\simeq 10$) in the generalized scenario. 

At $O(p^4)$ in the standard scheme \cite{GL83}, the amplitude depends
on four of the  LECs $l_i^r(\mu)$, the $N_f=2$ counterparts of the
$L_i^r(\mu)$. For $\mu\simeq M_\rho$, many observables turn out to be
dominated by chiral logs. This applies especially
to $a_0^0$ that increases from 0.16 to 0.20. This relatively big 
increase of 25$\%$ makes it necessary to go one step further in 
the chiral expansion.

The calculation of $O(p^6)$ was approached in two different ways. 
In the dispersive treatment \cite{KMSF95}, $A(s,t,u)$ was calculated
explicitly up to a crossing symmetric subtraction polynomial
\begin{equation}
[b_1 M_\pi^4 + b_2 M_\pi^2 s +b_3 s^2 +b_4 (t-u)^2]/F_\pi^4
+[b_5 s^3+b_6 s(t-u)^2]/F_\pi^6
\end{equation}
with six dimensionless subtraction constants $b_i$. Including
experimental information from $\pi\pi$ scattering at higher energies,
Knecht et al. \cite{KMSF96} evaluated four of those constants
($b_3$,\dots, $b_6$) from sum rules. 

The field theoretic calculation involving Feynman diagrams with
$L=0,1,2$ was performed in the standard scheme \cite{BCEGS96}.
Of course, the diagrammatic calculation reproduces
the analytically nontrivial part of the dispersive approach. 
Moreover, in the field theoretic approach the previous subtraction
constants are obtained as functions
\begin{equation}
b_i(M_\pi/F_\pi,M_\pi/\mu;l_i^r(\mu),r_i^r(\mu))~,
\end{equation}
where the $r_i^r$ are six combinations of LECs of the $SU(2)$
Lagrangian of $O(p^6)$ \cite{BCE2}.

Compared to the dispersive approach, the diagrammatic method offers
the following advantages:
\begin{description}
\item[i.]  The full infrared structure is exhibited to $O(p^6)$. In
particular, the $b_i$ contain chiral logs 
$l^n$ ($n\le 2$) that are known to be numerically
important, especially for the infrared dominated parameters $b_1$ and
$b_2$.
\item[ii.]  The explicit dependence on LECs makes
phenomenological determinations of these constants and comparison with
other processes possible. This is especially relevant for determining
$l_1^r$, $l_2^r$ to $O(p^6)$ accuracy.
\item[iii.]  The full dependence on the pion mass allows one to
evaluate the amplitude even at unphysical values of the quark mass.
One possible application is
to confront the CHPT amplitude with lattice calculations.
\end{description}

\begin{figure}[ht] 
\begin{center} 
\leavevmode 
\includegraphics[angle=-90,width=10cm]{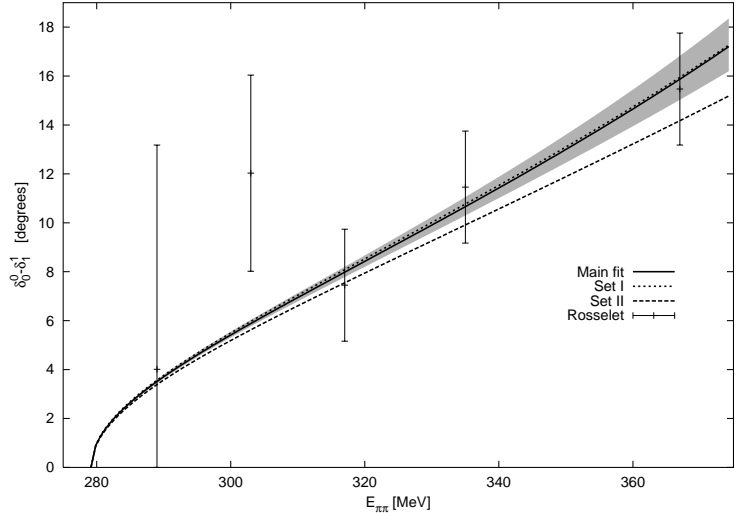}
\caption{Phase shift difference $\delta_0^0-\delta_1^1$ to $O(p^6)$
for different values of the $l_i^r(\mu)$ (Sets I, II 
\protect\cite{BCEGS96}; main fit \protect\cite{ABT00}) in comparison 
with experiment \protect\cite{Ross77}.}
\label{fig:ABT}
\end{center} 
\end{figure}

The original analysis \cite{BCEGS96} estimated the $O(p^6)$ LECs 
$r_i^r(\mu)$ from meson resonance exchange and gave results for two
different sets of the $l_i^r(\mu)$ ($i=1,\dots,4$). Not surprisingly, 
it turns out that the low partial waves are mainly sensitive to the 
LECs of $O(p^4)$. More recently, Amoros et al. \cite{ABT00} have
determined (some of) those LECs using $O(p^6)$ results whenever
available. The phase shift difference $\delta_0^0 - \delta_1^1$ that
can be extracted from $K_{e4}$ data is shown in Fig.~\ref{fig:ABT}
\cite{ABT00} together with experimental results from 1977 \cite{Ross77}.

Preliminary new results from the $K_{e4}$ experiment in Brookhaven
(BNL-E865 \cite{BNL865}) are in agreement with the theoretical
predictions. 

\subsection{Dispersive matching}
The most important development in recent years in the field of 
$\pi\pi$ scattering 
is the new dispersive analysis \cite{ACGL00} via Roy equations
\cite{Roy71} and the subsequent matching with the chiral amplitude 
\cite{CGL00} (that actually appeared after this meeting).

\begin{figure}[ht] 
\leavevmode \begin{center}
\includegraphics[angle=-90,width=10cm]{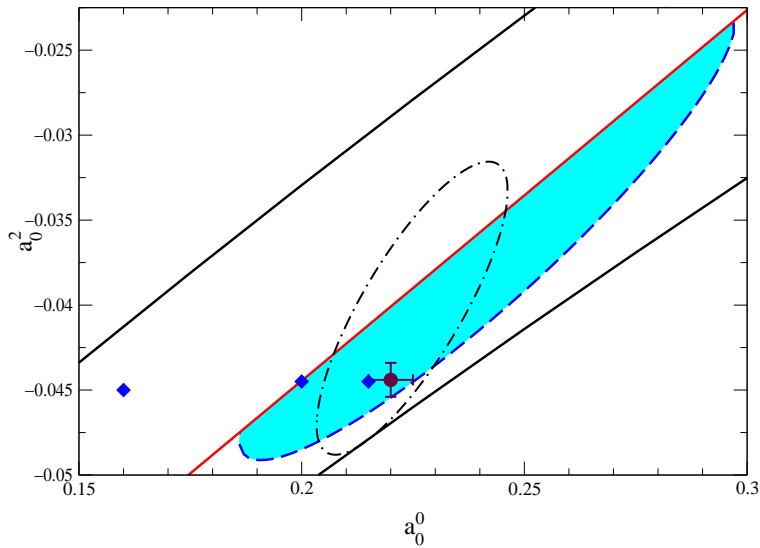}
\caption{Various constraints on the scattering lengths 
$a_0^0, a_0^2$ from Ref.~\protect\cite{CGL00}. The diamonds illustrate
the convergence of the CHPT results of orders $p^2, p^4$ and $p^6$.
The cross shows the final result of Ref.~\protect\cite{CGL00}
reproduced in (\protect\ref{eq:a02}).}
\label{fig:Roy}
\end{center}
\end{figure}

In a first step \cite{ACGL00}, the low partial waves ($S,P$) were
derived from dispersion relations (Roy equations) with high-energy
data ($E\ge 0.8$ GeV) as experimental input and with the
scattering lengths $a_0^0, a_0^2$ as subtraction constants. The
output comes in the form of amazingly precise predictions for the
phase shifts and for the remaining threshold parameters in terms of
$a_0^0, a_0^2$. However, even with the new results from BNL-E865
\cite{BNL865} included (the dash-dotted ellipse in Fig.~\ref{fig:Roy}),
the allowed domain in the $a_0^0-a_0^2$ plane shown in Fig.~\ref{fig:Roy}
is still rather large although previously admissible values for
$a_0^0$ of 0.26 or higher are practically ruled out now.

In a second step, Colangelo, Gasser and Leutwyler \cite{CGL00} have
matched the Roy-equation and $O(p^6)$ solutions at the unphysical point
$s=0$. The reason for this choice is that the chiral expansion
converges better at $s=0$ than at the physical threshold 
$s=4 M_\pi^2$. With the input of $l_3^r,l_4^r$ (main sensitivity) and
the $O(p^6)$ LECs $r_i^r$ ($i=1,\dots,4$), this matching produces
values for $l_1^r,l_2^r$, $r_5^r,r_6^r$ and especially for 
$a_0^0, a_0^2$.

Some of the resulting threshold parameters are collected in the last
column of Table \ref{tab:Roy}. For comparison, the original predictions
based on set I \cite{BCEGS96} of $O(p^4)$ LECs and the predictions
from the recent $O(p^6)$ analysis \cite{ABT00} are also shown.

\renewcommand{\arraystretch}{1.1}
\begin{table}[ht]
\begin{center}
\caption{Threshold parameters for $\pi\pi$ scattering}
\label{tab:Roy}
\vspace*{1cm}
\begin{tabular}{|c||c|c|c|}\hline
& & &   \\
       & $O(p^6)$ main fit \protect\cite{ABT00} & $O(p^6)$ 
set I \protect\cite{BCEGS96} &  matching solution 
\protect\cite{ACGL00,CGL00} \\ 
& & &  \\
\hline
$a^0_0$     &$0.219\pm0.005$ &$0.222$ & $0.220\pm 0.005$\\
$-10 a^2_0$ & $0.420\pm0.010$  & $0.420$ 
& $0.444\pm 0.010$\\
$b^0_0$     &$0.279\pm0.011$ &$0.282$ &$0.276\pm 0.006$\\
$-10 b^2_0$ & $0.756\pm0.021$ & $0.729$ 
&$0.803\pm 0.012$\\
$10 a^1_1$  &$0.378\pm0.021$ & $0.404$  &$0.379\pm 0.005$\\
$10^2 b^1_1$&$0.59\pm0.12$   & $0.83 $  &$0.567\pm 0.013$\\
$10^2 a^0_2$&$0.22\pm0.04$   & $0.28 $  &$0.175\pm 0.003$\\
$10^3 a^2_2$& $0.29\pm0.10$    & $0.24 $  
&$0.170\pm 0.013$\\
\hline
\end{tabular}
\end{center}
\end{table}

The following conclusions emerge concerning our present understanding
of $\pi\pi$ scattering in QCD.
 
\begin{description}
\item[i.] The chiral expansion ``converges''. There are no signs of
any unexpected large higher-order contributions.
\item[ii.] The biggest contributions in many cases are well
understood: they are due to chiral logs, especially for $S$-waves.
\item[iii.] The state-of-the-art amplitude arising from the
combination of CHPT to $O(p^6)$ and dispersion theory (Roy equations)
is in agreement with all available data. Additional forthcoming
experimental results should put the theory to even more stringent
tests. 
\item[iv.] Standard CHPT produces the following solid predictions
for the $S$-wave scattering lengths \cite{CGL00} :
\begin{eqnarray}  
a_0^0 &=& 0.220 \pm 0.005  \nl
a_0^2 &=& - 0.0444 \pm 0.0010~.
\label{eq:a02}
\end{eqnarray}
\item[v.] These results depend on the standard values of $l_3^r,l_4^r$
which Generalized CHPT may and does question. However, $a_0^0 < 0.25$
is a prediction from Roy equations only and therefore independent of 
the chiral expansion scheme.
\item[vi.] A second caveat concerns the precision reached by the
dispersive analysis. As stressed by the authors of Ref.~\cite{ACGL00},
the errors in the last column of
Table \ref{tab:Roy} should be interpreted as being due
to the ``experimental noise'' seen in the analysis. At this level of
accuracy, isospin violation and electromagnetic corrections must be
included. The latter are partly available for $\pi\pi \to \pi\pi$ 
\cite{KU98} but not (yet) for $K_{e4}$.
\end{description}

\setcounter{equation}{0}
\setcounter{subsection}{0}
 
\section{Isospin Violation}
\label{sec:iso}
There are two sources of isospin violation in the standard model:
\begin{itemize} 
\item[a.] $m_u \neq m_d$ (strong isospin violation);
\item[b.] Electroweak interactions, in particular
electromagnetic corrections.
\end{itemize} 

The level of accuracy reached in many processes, experimentally and/or
theoretically, calls for inclusion of the dominant isospin-violating
effects. As an example, consider the dependence of the $\pi\pi$
scattering length $a_0^0$ on the pion mass in lowest order:
\begin{equation} 
a_0^0 = \frac{7 M_\pi^2}{32 \pi F_\pi^2} = \left\{ \ba{ll}
0.159 & \quad\qquad M_\pi=M_{\pi^+} \\
0.149 & \quad\qquad M_\pi=M_{\pi^0} \ea \right. ~.
\end{equation} 
The difference is comparable to the corrections of $O(p^6)$.

To locate the leading isospin-violating effects in $\pi\pi$
scattering, let us first look at the relevant part of the strong
Lagrangian (for $N_f=2$ \cite{GL84}):
\begin{eqnarray}
\cL_2 + \cL_4  &=& \frac{F^2}{4} \langle D_\mu U 
D^\mu U^\dagger + 
\chi U^\dagger + \chi^\dagger  U \rangle \nl
&+& \dots - \frac{l_7}{16} \langle \chi U^\dagger - 
\chi^\dagger U \rangle^2 ~.
\end{eqnarray} 
The term proportional to $l_7$ makes a tiny contribution to
$M_{\pi^0}^2$ but there are no other consequences of $m_u \neq m_d$
for $\pi\pi \to \pi\pi$ to $O(p^4)$. The leading effect is of 
electromagnetic origin due to the one-parameter Lagrangian 
$\cL_{e^2 p^0}^{\rm em}$ in Table \ref{tab:EFTSM}:
\begin{equation} 
\cL_{e^2 p^0}^{\rm em} = e^2 Z F^4 \langle Q U Q U^\dagger \rangle
\label{eq:e2p0}
\end{equation} 
\begin{equation}  
\to \qquad \Delta M_\pi^2 = M_{\pi^+}^2
- M_{\pi^0}^2 = 2 e^2 Z F^2 ~, \quad Z \simeq 0.8 ~.
\label{eq:mdiff}
\end{equation} 
Again, there are no other contributions to $\pi\pi$ scattering in
addition to the kinematical effect due to the mass shift 
(\ref{eq:mdiff}). The
genuine leading electromagnetic (and isospin-violating) corrections
to the scattering amplitude are of $O(e^2 p^2)$. 

\subsection{Pionic atoms}
The DIRAC experiment \cite{Dirac00} at CERN attempts to measure the
lifetime of electromagnetically bound $\pi^+-\pi^-$ ``atoms'' in the
ground state to 10$\%$ accuracy. The width is dominated by the
transition $\pi^+ \pi^- \to \pi^0 \pi^0$, with the $\pi^0$'s decaying
subsequently into photons. Assuming isospin invariance for the
amplitude, the decay width into two neutral pions is given by 
(LO stands for lowest order, i.e., without isospin violation)
\cite{DGBT54}
\begin{equation} 
\Gamma_{2\pi^0}^{\rm LO} = \frac{2}{9}\alpha^3 p^* (a_0^0-a_0^2)^2 ~\to~
\tau \simeq 3\times 10^{-15}s
\end{equation} 
where $p^* = (M_{\pi^+}^2-M_{\pi^0}^2 -M_{\pi^+}^2
\alpha^2/4)^{1/2}$ is the three-momentum of either $\pi^0$ in the final
state (in the center of mass) and $a_0^0-a_0^2$ is the scattering
length for $\pi^+ \pi^- \to \pi^0 \pi^0$. Therefore, DIRAC is expected 
to measure $|a_0^0-a_0^2|$ to 5$\%$ accuracy (cf. Fig.~\ref{fig:Roy}).  

As already emphasized, this level of accuracy makes a careful estimate
of isospin-violating corrections mandatory. Since in QCD $A(\pi\pi \to
\pi\pi)$ does not contain terms linear in $m_u-m_d$ the leading
corrections are $O(\delta)$ with $\delta=\alpha$ or $(m_u-m_d)^2$
\cite{GGLR99}.

The effective field theory technique of Gall et al. \cite{GGLR99} for 
calculating those
corrections turns out to be superior to other previously or currently
employed methods, like nonrelativistic potential or Bethe-Salpeter
approaches. The method consists of the following main steps:
\begin{itemize} 
\item The starting point is CHPT for $N_f=2$ with a dynamical photon.
\item From CHPT, one passes to a nonrelativistic effective Lagrangian,
similar to the procedure employed in QED \cite{CL86}. To the order
required, the electromagnetic interaction occurs only through the
Coulomb potential \cite{GGLR99}.
\item With this nonrelativistic Lagrangian, one calculates the energy
of the bound state and the width $\Gamma_{2\pi^0}$ and matches the
scattering amplitude to the full relativistic expression with
$O(\delta)$ corrections included.
\end{itemize} 

The final result \cite{GGLR99} is a corrected rate
\begin{equation} 
\Gamma_{2\pi^0} = \frac{2}{9}\alpha^3 p^* A^2(1+K)
\end{equation} 
where $A$ is directly related to the relativistic on-shell scattering
amplitude at threshold\footnote{The symbol $o(\delta)$ stands for
terms vanishing faster than $\delta$.}:
\begin{equation} 
A= -\frac{3}{32\pi} \RE A_{\rm thr}(\pi^+\pi^-\to
\pi^0\pi^0) + o(\delta)~.
\end{equation}
The quantity $K$ contains additional corrections :
\begin{eqnarray}
K &=& \frac{\Delta M_\pi^2}{9 M_\pi^2}(a_0^0+2 a_0^2)^2 
- \frac{2\alpha}{3}(\ln \alpha -1)(2 a_0^0+a_0^2)+o(\delta) \nl
&=& 1.1 \times 10^{-2}~.
\end{eqnarray}
The threshold amplitude $A$ is expanded to first order in $\delta$ :
\begin{equation} 
A=a_0^0 - a_0^2 + h_1(m_u-m_d)^2 +h_2 \alpha + o(\delta)~.
\end{equation}
The scattering lengths $a_0^0,a_0^2$ and the coefficients $h_i$ in 
this expansion are to be taken in the
isospin limit for $M_\pi=M_{\pi^+}$. Since $h_1$ turns out to be
negligible the uncertainty in the final result \cite{GLR99}
is exclusively due to
$h_2$ that depends both on the LECs $l_i^r$ (multiplied by the
electromagnetic coupling constant $Z$ defined in Eq.~(\ref{eq:e2p0})) and
on LECs of $O(e^2p^2)$ \cite{Urech95}:
\begin{equation} 
A=a_0^0 - a_0^2 + \ve ~, \qquad \ve=(0.58 \pm 0.16)\times
10^{-2} ~.
\end{equation}

Putting everything together, Gasser et al. \cite{GLR99} obtain for the
relative isospin-violating correction 
\begin{equation} 
\displaystyle\frac{\Gamma_{2\pi^0}-
\Gamma_{2\pi^0}^{\rm LO}}{\Gamma_{2\pi^0}^{\rm LO}}=
\underbrace{\displaystyle\frac{2\ve}{a_0^0-a_0^2}}_
{0.047}+ \underbrace{K}_{0.011}=0.058 ~.
\end{equation} 
Since the experimental accuracy for $\Gamma_{2\pi^0}$ is expected to
be 10$\%$ the above 6$\%$ correction due to isospin violation is 
crucial for extracting $|a_0^0 - a_0^2|$ from
experiment. This combination can then be directly compared to the
CHPT calculation for $\alpha=0, m_u=m_d$, as discussed in
Sec.~\ref{sec:pipi}.

\subsection{CP violation and $\pi^0-\eta$ mixing}
Isospin violation is usually a small effect. Unlike for $\pi\pi$
scattering where terms linear in $m_u-m_d$ are absent, the following
order-of-magnitude estimates should be relevant for $N_f=3$, in particular
for $K$ decays:

\begin{center} 
\begin{tabular}{c|c}
$O(m_u-m_d)$ & $O(\alpha)$ \\
\hline \\
$\displaystyle\frac{M_{K^0}^2 - M_{K^+}^2}{M_K^2} \sim 1.5\%$ & 
$\displaystyle\frac{M_{\pi^+}^2 - M_{\pi^0}^2}{M_K^2} \sim 0.5\%$   
\end{tabular}
\end{center} 

In general, strong isospin violation and electromagnetic corrections are
comparable in size and the effect is quite small unless there is some specific
enhancement mechanism. This is precisely the case for the dominant
decays of kaons into two pions. Restricting the discussion to $K^0$
decays, the amplitudes in the isospin limit are usually parametrized as
\cite{HBB}
\begin{eqnarray}  
A(K^0 \to \pi^+ \pi^-) &=& A_0 e^{i\delta^0_0}+
\frac{1}{\sqrt{2}} A_2 e^{i\delta^2_0} \nl 
A(K^0 \to \pi^0 \pi^0) &=& A_0 e^{i\delta^0_0}-
\sqrt{2} A_2 e^{i\delta^2_0} 
\end{eqnarray}
with $\delta_{l=0}^I$ the  $\pi\pi$ phase shifts at $s=M_K^2$.
Although the amplitudes $A_0, A_2$ could well be similar in size (in
fact, they are in the so-called ``naive'' factorization limit) they
are actually quite different in the real world as expressed by the
$\Delta I=1/2$ ``rule'':
\begin{equation} 
A_2/A_0 \simeq 1/22 ~.
\end{equation}
This rule suggests a possible strong enhancement of 
isospin-violating corrections in $A_2$, e.g.,
\begin{equation} 
A_2^{\rm ind}/A_2 \sim \frac{M_{K^0}^2 - M_{K^+}^2}{M_K^2} \cdot
\frac{A_0}{A_2} \simeq 0.35 ~.
\end{equation} 
One instance where this enhancement can be seen is the CP violating
ratio $\ve'/\ve$. In the approximate formula \cite{bosch00}
\begin{equation}
\frac{\ve'}{\ve} \approx 13\,\IM V_{ts}^* V_{td} 
\left[ B_6^{(1/2)}(1-\Omega_{IB}) - 0.4 B_8^{(3/2)}\right]
\end{equation}
the quantity
\begin{equation} 
\Omega_{IB}:=\frac{\IM A_2^{\rm IB}\RE A_0}{\IM A_0 \RE A_2}
\end{equation}
is a measure of strong isospin violation. The
so-called bag factors \cite{HBB} $B_6^{(1/2)},B_8^{(3/2)}$ are
parameters of $O(1)$ and $V_{ij}$ are Cabibbo-Kobayashi-Maskawa 
matrix elements.

To lowest order in the chiral expansion, $\Omega_{IB}$ is due to
$\pi^0-\eta$ mixing:
\begin{eqnarray}  
\Omega_{IB} &=& \frac{2 \sqrt{2} \epe^{(2)}}{3 \sqrt{3} 
\omega} = 0.13 \\
\epe^{(2)} &=& \frac{\sqrt{3}(m_d - m_u)}{4 (m_s - \hat m)}~.
\end{eqnarray} 
At $O(p^4)=O[(m_u-m_d)p^2]$, there is a large contribution from
$\eta^\prime$ exchange \cite{etaetap}:
\begin{equation} 
\Omega_{\eta + \eta'} = 0.25 \pm 0.08 ~.
\label{eq:etap}
\end{equation}
Such a big correction raises the legitimate question
whether there are other contributions to $\Omega_{IB}$ at the same 
order. One part of the problem is related to the higher-order
corrections to the $\pi^0-\eta$ mixing angle. Writing
$$
\epe = \epe^{(2)} + \epe^{(4)} ~,
$$
one finds \cite{GL85b,EMNP00}
\begin{eqnarray} 
\epe^{(4)}/\epe^{(2)} &=& \frac{128 (M^2_K -
M^2_\pi)^2}{3 F_\pi^2 (M_\pi^2 - M_\eta^2)} \left[3 L_7 +
L^r_8(M_\rho)\right] \nl
&+& {\rm ~(small)~~ loop ~~corrections}~.
\label{eq:eps4}
\end{eqnarray}
Of the three contributions in (\ref{eq:eps4}), the first one contains
the previously mentioned $\eta^\prime$ exchange:
\begin{equation} 
\epe^{(4)}(L_7)/\epe^{(2)}= 1.10~,
\end{equation} 
perfectly consistent with the old estimate (\ref{eq:etap}). The
surprise comes from the second term that was not included previously
and in fact almost cancels the $L_7$ term \cite{EMNP00}:
\begin{equation}
\epe^{(4)}(L_8^r(M_\rho))/\epe^{(2)}= - 0.83 ~.
\end{equation}
One reason why this part was not included originally is the physical
interpretation of the LEC $L_8$: it is dominated by 
$a_0(980)$ exchange and its relevance for $\Omega_{IB}$ is 
much less transparent than $\eta^\prime$ exchange. The loop
corrections in (\ref{eq:eps4}) are negligible for $\mu=M_\rho$.

The total contribution from $\pi^0-\eta$ mixing to $\Omega_{IB}$ 
with
$$
3 L_7 + L_8^r(M_\rho)= (-0.25 \pm 0.25) \cdot 10^{-3}
$$
amounts to
\begin{equation} 
\Omega_{IB}^{\pi^0\eta} = 0.16 \pm 0.03 ~.
\label{eq:omfinal}
\end{equation} 
With the preferred  bag factors of the Munich group \cite{bosch00},
a decrease from $\Omega_{IB}=0.25$ to 0.16 implies an increase of
$\ve'/\ve$ by 21$\%$. 

The systematic analysis of isospin-violating corrections at
next-to-leading order also shows that there are additional
contributions proportional to $m_u-m_d$ that depend on largely unknown
weak LECs. Model-dependent estimates suggest \cite{IBLEC} an
additional decrease of $\Omega_{IB}$ below (\ref{eq:omfinal}).

\setcounter{equation}{0}
\setcounter{subsection}{0}
 
\section{Outlook}
\label{sec:summ}
Chiral perturbation theory is the effective field theory of the
standard model at low energies. It is a nonrenormalizable but
perfectly well-defined and respectable quantum field theory. One of
its major advantages is the absence of double counting: in contrast to
many phenomenological models, only hadronic fields enter in the
chiral Lagrangians. The price for the generality of CHPT is the 
abundance of low-energy constants at higher orders in the chiral
expansion. 

CHPT has undoubtedly reached a level of maturity. In the meson sector,
two-loop calculations are state of the art for the strong
interactions and there is little to gain from going even further. The
main task in this area for the near future  is the evaluation of
``small'' effects due to isospin violation and electromagnetic
corrections. There is more work to be done for the nonleptonic weak
interactions \cite{HBB}.

For single-baryon processes, even the complete one-loop amplitudes 
remain to be calculated for some transitions.
The slow convergence of the chiral expansion may
require a reordering of the series. Chiral symmetry in
nuclear physics is still a rapidly expanding field.

Even a superficial look at the effective Lagrangian of the standard
model in Table \ref{tab:EFTSM} tells us that phenomenology cannot
possibly fix all the low-energy constants of this Lagrangian. Progress
in the field will depend to a large degree on the success of
supplementary methods to determine those parameters.


\section*{Acknowledgements}
\noindent
I thank Santi Peris, Vicente Vento and their staff for having
organized this School in such beautiful surroundings.


\end{document}